# Super-Bandgap Electroluminescence from Cesium Lead Bromide

JUSTIN SCULLEY,[1] JEREMY KOWKABANY,[1] DIANA K. LAFOLLETTE,[2] CARLO PERINI,[2] YAN XIN,[3] JUAN-PABLO CORREA-BAENA,[2] AND HANWEI GAO[1]*

1. Department of Physics, Florida State University, Tallahassee, FL 32306

2. School of Materials Science and Engineering, Georgia Institute of Technology, Atlanta, GA 30332

3. National High Magnetic Field Laboratory, Tallahassee, FL 32310

ABSTRACT Halide perovskites is a new class of semiconductors with exceptional optoelectronic properties. Among many advantages offered by halide perovskites, the bandgap energy can be tuned in a much broader range than what was possible in conventional semiconductors. This was commonly achieved in previous research by mixing different species of halides into solid solutions. The tuned bandgap using this method, however, often underwent an energy shift under optical or electrical stimuli due to halide segregation. In this work, we discovered an alternative approach to achieve super-bandgap electroluminescence from $CsPbBr_3$. The peak energy of the light emission can be 0.7 eV higher than the reported bandgap energy. Evidence pointed to the radiative recombination at the perovskite-PEDOT:PSS interface being responsible for the unexpected blueshift of electroluminescence. We speculated that perovskite nanocrystals were formed therein and produced higher-energy photons due to quantum confinement. The results suggested an alternative strategy to manipulate and stabilize the color of electroluminescence and achieve particularly blue emission in perovskite-based LEDs.

**Introduction:**

Halide perovskites have shown tremendous potential for optoelectronic applications. Efficient solar cells[1] and luminous light emitting diodes (LEDs),[2] for example, have been successfully demonstrated in research laboratories. Among many advantages, the bandgap energy of the perovskites can be tuned by alloying different halide species into homogeneous crystals. Without a miscibility gap, the halides can be mixed with virtually arbitrary atomic ratios.[3] As a result, the perovskites offer much greater tunability than that achievable in traditional semiconductors (e.g. III-Vs or II-VIs), covering the entire visible and a part of the ultra-violet and near-infrared spectrum.[4] For lighting and display applications, bandgap energy determines the color of light emission. The capability of broadband tuning is, therefore, essential to access the full calumet space for the best color rendering performance.

Bandgap tuning in the perovskites, however, is not perfect. Because of the inherent ion migration and unique thermodynamics under external stimuli, homogenous mixed-halide perovskites often underwent phase segregation upon optical illumination or electrical biases.[5-8] While the underlying mechanism remains debatable, the phase segregation was evident by both optical and crystallographic characteristics.[9] The resulting shift of bandgap energy limited the applicability of the bandgap tuning in practical optoelectronic devices.[10, 11]

In this work, we implemented LED devices using $CsPbBr_3$, a type of perovskite emitter with all-inorganic composition. Surprisingly, the photon energy of electroluminescence could be substantially higher than the expected bandgap energy. We ascribed the apparent super-bandgap emission to the formation of quantum-confined nanocrystals near the interface against the polymeric hole transport layer. By adjusting the charge injection balance, radiative recombination could be precisely restricted to the nanocrystal region. This phenomenon suggested an alternative strategy to manipulate the wavelength of emission in perovskite-based LEDs. Without involving mixed-halide perovskites, the previous problem with phase segregation and bandgap shift can be circumvented. Electroluminescence with stable spectral output is therefore expected using this approach.

**Results and discussion:**

Our devices were constructed following the previously reported p-i-n architecture (**Fig. 1**).[12] PEDOT:PSS and TPBi were used as hole and electron injection materials, respectively. A layer of solution-processed $CsPbBr_3$ was sandwiched in-between, serving as the light emitter. The interface between the TPBi and aluminum cathode was functionalized with a very thin layer of LiF, reducing the electrical potential barrier and facilitating efficient electron injection. The thicknesses of the TPBi (p), perovskite (i) and PEDOT (n) layers were approximately 40 nm, 110 nm and 135 nm, respectively. More details of device fabrication are described in Methods.

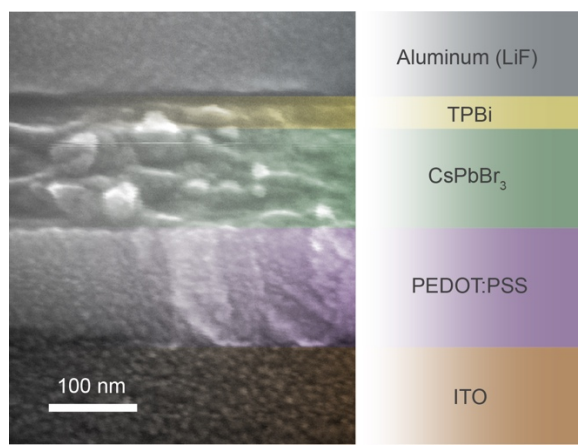

**Figure 1. A cross-sectional SEM image shows the layered architecture of an LED device.**

The devices exhibited diode behaviors typical of perovskite-based LEDs. The electrical turn-on was observed around 2.4 volts, whereas the luminance started to rise abruptly past 2.9 volts (**Fig. 2A**). A maximal external quantum efficiency (EQE) of 0.2% was reached with a current density of 150 mA/cm$^2$, beyond which the efficiency decreased gradually (**Fig. 2B**). Such so-called roll-over behavior was often attributed to imbalance in charge injection[13] or quenching of luminescence due to Auger recombination.[14, 15] While the scope of this work was not about pursuing the best device performance, we believe there was substantial room for improvement in both electrical and luminescent efficiencies. Particularly, the (pre-turn-on) leakage current in our devices was considerably high. Such phenomenon was often associated with ineffective charge confinement within the emissive layer, which could be offset by increasing the thickness of the emissive layer or introducing electron- or hole-blocking layers.[16] The high leakage current also limited the current-to-photon yield, partially responsible for the limited luminosity and EQE observed.

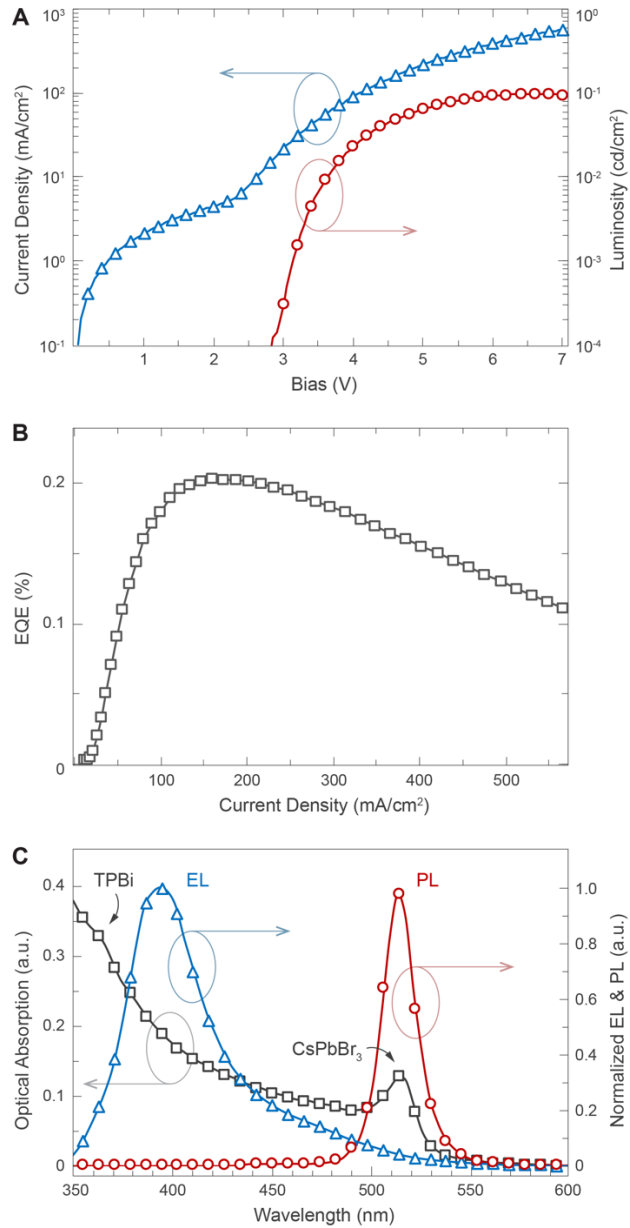

**Figure 2. Device characteristics of blue-emitting LEDs. (A)** Clear turn-ons can be observed in both IV and LV relations. **(B)** The device reached a maximum EQE of 0.2 % with 150 mA/cm2 current density. **(C)** While both PL and optical absorption spectra showed the excitonic peak near the bandgap energy of $CsPbBr_3$, the EL was peaked at a substantially shorter wavelength (higher energy).

What was the most striking was the large spectral shift between photoluminescence and electroluminescence. The photoluminescence spectrum was dominated by a single, narrow peak centered around 517 nm (**Fig. 2C**). This wavelength was consistent with the well-recognized exciton energy near the bandgap in $CsPbBr_3$.[17] The excitonic feature was also observed in the optical absorption spectrum. While an absorption edge would be typically expected at the bandgap energy of direct-bandgap semiconductors, the sizable exciton binding energy of the halide perovskite made the excitonic features more outstanding as a spectral peak. Note that the small shoulder near 365 nm can be attributed to the optical absorption of TPBi,[18, 19] provided that the spectrum was measured from a completed device with all the transport layers stacked.

While the photoluminescence matched the bandgap energy, the electroluminescence was observed at much shorter wavelengths, with a peak wavelength at 395 nm (**Fig. 2C**). The blue/purple electroluminescence could not be generated in the bulk of the $CsPbBr_3$ layer, given the large (> 0.7 eV) difference between the blue edge of the electroluminescence and the bandgap energy of $CsPbBr_3$. TPBi was unlikely the emitter either, as the maximum luminance and EQE observed here largely exceeded the luminescent yield of TPBi reported previously. Our hypothesis of the mechanism behind this super-bandgap luminescence can be found later in the discussion.

The wavelength of electroluminescence was found to be tunable. As the thickness of the hole transport layer (PEDOT:PSS) was reduced, the color of electroluminescence shifted from blue/purple to green (**Fig. 3**). The shift was not continuous in the wavelength domain. Instead, a new peak emerged around 520 nm (green emission), accompanied by a decrease in intensity of the blue/purple emission. When the PEDOT layer was sufficiently thin (<45 nm), the electroluminescence became purely green. The spectrum under this condition matched well with the photoluminescence of the devices (Fig. 1C), corresponding to the excitonic recombination that one would normally expect near the bandgap energy of CsPbBr$_3$.

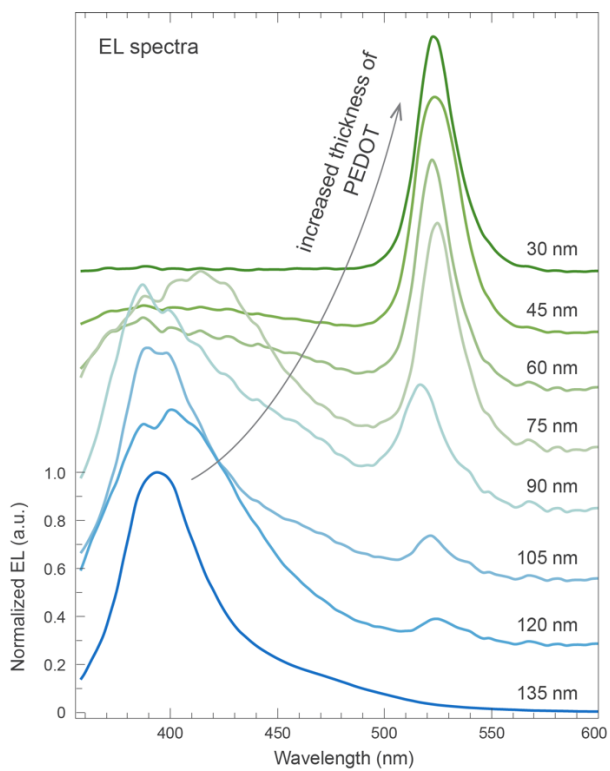

**Figure 3. The wavelength of electroluminescence was tunable.** By increasing the thickness of hole transport layer (PEDOT:PSS), the peak wavelength of electroluminescence shifted from 395 nm to 520 nm.

The correlation between the electroluminescence spectra and the thickness of PEDOT indicated that the balance of injected charge carriers could be essential to the super-bandgap emission observed. In a simplified model, charge injection rates are inversely proportional to the thickness of the transport layer.[20, 21] With a thinner electron transport layer, for instance, the electron injection would experience lower electrical resistance, resulting in a higher electrons injection rate.

The varied balance between the injected electrons and holes could shift the zone of radiative recombination in the emissive layer.

In our devices, specifically, the blue/purple electroluminescence was observed when the (hole transport) PEDOT layer was relatively thick. The hole-deficient (electron-rich) condition of had to shift the radiative recombination zone closer to the PEDOT-perovskite interface. Alternatively, when a thinner PEDOT layer was incorporated, the recombination zone would be pushed further away from that interface and more into the bulk of the perovskite film. The fact that the long-wavelength EL (green) emerged quickly as the PEDOT layer was thinned suggested that these "blue emitters" ought to be highly confined to that interface. The thickness of this interfacial layer had to be so small that the contribution from the blue emitters was largely negligible in photoluminescence spectra. Beyond the territory of the interface, the perovskite layer consisted primarily of normal, green-emitting $CsPbBr_3$. That explained why the photoluminescence spectra remained nearly identical, regardless of the electroluminescence being blue/purple or green. Note that the thickness of perovskite layer was on the same order of magnitude as the optical penetration depth deduced from the optical absorbance.[22] It was then reasonable to assume that the photoluminescence was primarily contributed by the bulk of the perovskite layer, but not bound to any interface.

While the exact mechanism of blue EL remained elusive, one of the possibilities could be the formation of perovskite nanocrystals near the perovskite-PEDOT interface (**Fig. 4A**). When the size of semiconducting nanocrystals is sufficiently small, the optical bandgap energy would increase – an effect known as quantum confinement. The amount of spectral shift of radiative recombination is determined by the size of nanocrystals and the exciton Bohr radius. For $CsPbBr_3$, the nanocrystals had to be smaller than 3 nm to generate the blue-shifted EL observed around 400 nm.[23] In our devices, the nanocrystals must be located at the proximity of the PEDOT-perovskite interface to be consistent with the electroluminescence as a function of the PEDOT thickness discussed earlier. The broad spectral width of the EL spectra suggested that the size of the nanocrystals was be monodispersed, but rather distributed across a wide range.

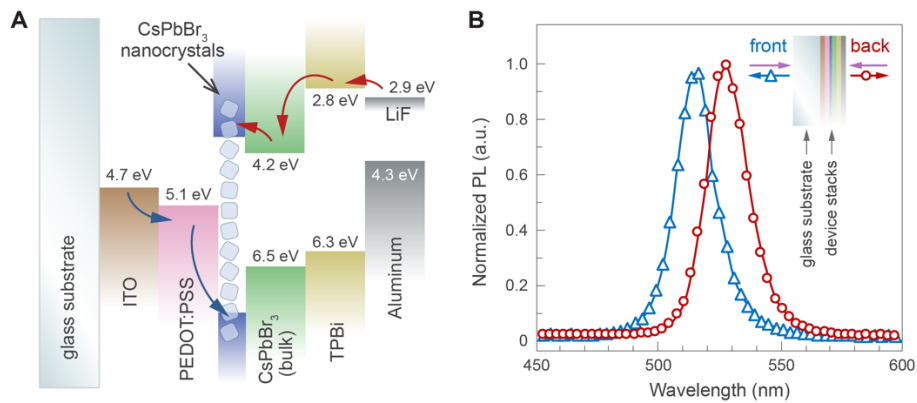

**Figure 4. A proposed mechanism for the observed super-bandgap electroluminescence. (A)** A schematic of the device band alignment illustrates the electroluminescence from quantum-confined nanocrystals formed at the perovskite-PEDOT interface. **(B)** Photoluminescence measured from front and back of the devices showed a spectral shift.

The formation of perovskite nanocrystals could be templated by nanoscale pores at the surface of PEDOT:PSS. When perovskites were synthesized on top of PEDOT:PSS, the precursor solution got infiltrated into the nanopores and crystalized there into small nanoparticles. Unfortunately, we were not able to visualize these nanocrystals using any electron microscopy techniques. The challenges could originate from the polycrystalline morphology of the perovskite layer and the extremely small thickness of the nanocrystal-containing interface. An interesting and perhaps relevant observation was the spectral shift between the photoluminescence measured from the front or the back of the devices (**Fig. 4B**). Although the difference between the blue- and green-emitting EL was considerably larger, this shift in PL at least showed the CsPbBr$_3$ near the PEDOT interface exhibited optical behavior different from that of the bulk of perovskite layer.

**Conclusions:**

The results of this work suggested an alternative strategy to manipulate the color of electroluminescence in perovskite-based LEDs. Compared to the most reported methods based on mixed-halide perovskites, our approach of wavelength tuning does not involve alloying halides and, therefore, would not suffer from the halide segregation and spectral shift during continuous device operation. Our method may also outperform the previous efforts using colloidal quantum dots to implement perovskite LEDs with wavelength tunability. To limit the crystal growth and control the size distribution, colloidal quantum dots were often synthesized with capping ligands. When incorporated in devices, these electrical insulating ligands inhibited charge injection into the quantum dots and therefore reduce the current-to-photon yield. In our devices, the formation of perovskite nanocrystals were likely templated by the porous structures of at the surface of the PEDOT:PSS layer. Charge injection would not be inhibited, if not enhanced, with such configuration provided that PEDOT:PSS itself functioned as the hole transport layer in the p-i-n junction. Besides LEDs, the nanocrystals templated using conductive polymers may also be incorporated into other perovskite-based optoelectronic devices such as solar cells or photodetectors for optimized spectrum of photoresponse.

**Methods:**

*Preparation of perovskite precursor solution*

Cesium bromide (CsBr) and lead bromide ($PbBr_2$) were dissolved with a molar ratio of 1:1 in dimethyl sulfoxide (DMSO), with a total concentration of 0.2 M. A second solution of 10 mg/ml polyethylene oxide (PEO) in DMSO was similarly prepared. Each solution was stirred separately in a sealed vial at 70 C for 2 hours and filtered with 0.22 mm PFTE filters. The $CsPbBr_3$ and PEO solutions were then combined at a volume ratio of 2:3 followed by another 2-hour stirring at 70 C to produce the final precursor solution.

*Materials synthesis and device fabrication*

Glass substrates coated with prepatterned indium tin oxide (ITO) were acquired from commercial sources. The surface of ITO was cleaned and prepared using oxygen plasma at a pressure of 4 Pa for 15 min. Solution of poly(3,4-ethylenedioxythiophene):polystyrene sulfonate (PEDOT:PSS) was spin-coated onto the ITO surface for 60 s. The spinning rate was varied between 1500 and 4000 rpm, corresponding to different thickness of PEDOT:PSS from 135 to 30 nm. The PEDOT:PSS layer was thermal-annealed and solidified in dry air in a tube furnace at a temperature of 150 C for 15 min. The samples were then transferred into a dry nitrogen glovebox for the remainder of the device fabrication.

In the glovebox, the precursor solution was spin-coated on top of the PEDOT:PSS at a speed of 3000 rpm for 60 s. The samples were covered immediately with a glass petri dish and annealed on a hot plate at 70 C for 10 min. To complete the LED devices, a thermal evaporator connected to the glovebox was used to deposit layers of 40-nm 1,3,5-tris(1-phenyl-1hbenzimidazol-2-yl)benzene (TPBi), 1-nm lithium fluoride (LiF) and 100-nm aluminum were evaporated sequentially on top of the perovskite layer.

*Measurements of optical and electrical properties*

Current density-voltage-luminance (J-V-L) characteristics were tested using two Keithley 2400 source meter units and a Thorlabs FDS1010 photodiode. Electroluminescence spectra was measured using an Ocean Optics SR2 spectrometer. Photoluminescence (PL) spectra was measured using a Nikon TE2000 inverted microscope coupled with a Horiba iHR-320 spectrometer. A HB-10103AF mercury lamp provided the optical excitation at ultraviolet wavelength. All measurements were taken in the N2 glove box at room temperature.

ASSOCIATED CONTENT

**Supporting Information**

AUTHOR INFORMATION

**Corresponding Author:** Hanwei Gao

E-mail: hanwei.gao@fsu.edu

**Author Contributions**

The research plans were conceived by both J.S. and H.G. J.S synthesized the materials and fabricated the devices. J.S and J.K carried out the electrical and optical measurements. Y.X. performed electron microscopy imaging and spectroscopy. D.K.L, C.P., and J.C. performed cathodoluminescece mapping and spectroscopy. J.S. and H.G drafted the manuscript and all authors edited the manuscript.

**Funding Sources**

National Science Foundation (2131610) and the Florida State University CRC SEED Award.

ACKNOWLEDGMENT

The authors appreciate insightful discussions with Prof. Letian Dou from Purdue University. The work was financially supported by National Science Foundation (award number 2131610). J.S. was partially funded by the InternFSU Program at Florida State University. H.G. acknowledges the support from the FSU CRC SEED Award.